\documentclass[aps,pra,pdf,superscriptaddress,twocolumn,showpacs,nofootinbib]{revtex4-2}

\usepackage{amsmath,amssymb,amsfonts}
\usepackage{eqnarray}
\usepackage[newcommands]{ragged2e}
\usepackage{graphicx,caption}
\usepackage{float}
\usepackage{subfig}
\usepackage{bm}
\usepackage{braket}
\usepackage{amsmath}
\usepackage{physics}
\usepackage{mathtools}
\usepackage{graphicx,color,xcolor,colortbl}
\usepackage{tikz}
\usepackage{microtype}
\usepackage[title]{appendix}
\usetikzlibrary{shapes}
\usepackage[colorlinks=true,
linkcolor=blue,
filecolor=magenta,      
urlcolor=blue,
citecolor=blue]{hyperref}

\definecolor{Gray}{gray}{0.85}
\definecolor{LightCyan}{rgb}{0.25,0.9,0.95}
\definecolor{LightMagenta}{rgb}{0.91,0.21,0.96}

\graphicspath{{images/}}
\makeatletter
\newcommand*{\rom}[1]{\expandafter\@slowromancap\romannumeral #1@}
\makeatother

\begin{document}
\title{{\bf Rayleigh-Taylor instability in a phase-separated three-component\\ Bose-Einstein condensate} }
\author{Arpana Saboo}%
\email{arpana.saboo@gmail.com}
\author{Soumyadeep Halder}
\author{Subrata Das}
\author {Sonjoy Majumder}
\email{sonjoym@phy.iitkgp.ac.in}
\affiliation{%
    Department of Physics, Indian Institute of Technology Kharagpur, Kharagpur, West Bengal 721302, India
}%

\begin{abstract}
    We investigate the Rayleigh-Taylor instability at the two interfaces in a phase-separated three-component Bose-Einstein condensate in the mean-field framework. The subsequent dynamics in the immiscible three-component condensate has been studied in detail for different cases of instigating the instability in the system. The rotational symmetry of the system breaks when the atom-atom interaction is tuned in such a way that the interface between the components becomes unstable giving rise to non-linear patterns of mushroom shapes which grow exponentially with time. We also identify these non-linear patterns as the solutions of the angular Mathieu equation, representing the normal modes. 
\end{abstract}
\maketitle
\section{Introduction}
\vspace{-0.2cm}
The \textit{Rayleigh-Taylor} instability (RTI) enfolds \cite{Rayleigh_1883,taylor_1997_instabilityliquidsurfaces,lewis_1997_instabilityliquidsurfaces,book_1961} at the interface between two fluids of different densities when the lighter fluid pushes the heavier one, making the system energetically unstable such that infinitesimal modulations begin to arise at the interface. This perturbation grows exponentially such that the two fluids tend to exchange their positions. However, for immiscible fluids sharing a flat interface, such an exchange is restricted without compromising the rotational symmetry of the system. This symmetry breaking leads to the deformation of the interface into complicated non-linear patterns including mushroom shapes \cite{daly_1967_numericalstudytwo}.
Thus, RTI is an interfacial instability which plays a vital role in a wide assortment of non-equilibrium phenomena occurring in nature ranging from laboratory to astronomical supernova explosions \cite{burrows_2000,bychkov_2006_dynamicsbubblessupernovae}, imploding targets in inertial confinement fusion \cite{sakagami_1990_threedimensionalrayleightaylorinstability}, collapsing cavitation bubbles \cite{plesset_1954,brenner_1995_bubbleshapeoscillations}, etc.
\par
Amongst superfluids, trapped dilute multi-component Bose-Einstein condensates (BECs) have been considered to be an exceptionally versatile test bed for theoretically investigating many novel interface phenomena \cite{kwon_2021_spontaneousformationstarshaped,bezett_2010_magneticrichtmyermeshkovinstability,kobyakov_2011_interfacedynamicstwocomponent,indekeu_2015_staticinterfacialproperties,indekeu_2018_capillarywavedynamicsinterface,kasamatsu_2001_macroscopicquantumtunneling}.
 The ability to tune the atom-atom interactions in the multi-component BECs through Feshbach resonance  offers an excellent platform to study the dynamics of coherently coupled mixtures. \cite{myatt_1997_productiontwooverlapping,hall_1998_dynamicscomponentseparation,miesner_1999_observationmetastablestates,mertes_2007_nonequilibriumdynamicssuperfluid,eto_2015_suppressionrelativeflow,modugno_2002_twoatomicspecies}.  Various aspects of two-component BECs including both static and dynamic facets of phase-separated BECs \cite{ho_1996_binarymixturesbose,pu_1998_propertiestwospeciesbose,timmermans_1998_phaseseparationboseeinstein,mazets_2002_wavesinterfacetwo,indekeu_2004_extraordinarywettingphase,kasamatsu_2009_vortexsheetrotating,lee_2016_phaseseparationdynamics}, such as interfacial instabilities and pattern formations, have been thoroughly investigated in literature \cite{maity_2020_parametricallyexcitedstarshaped,kobyakov_2012_parametricresonancecapillary,ao_1998_binaryboseeinsteincondensate,kasamatsu_2004_multipledomainformation,raju_2005_modulationalinstabilitytwocomponent,ronen_2008_dynamicalpatternformation,graham_1998_collectiveexcitationstrapped,pu_1998_collectiveexcitationsmetastability,gordon_1998_excitationspectruminstability,svidzinsky_2003_normalmodesstability,coen_2001_domainwallsolitons,bezett_2010_magneticrichtmyermeshkovinstability}. The RTI has been theoretically studied for trapped two-component BECs for the formation of mushroom-like patterns \cite{burmistrov_2009_rayleightaylorinstabilitycrystallization,Cabot_2006,sasaki_2009_rayleightaylorinstabilitymushroompattern, gautam_2010_rayleightaylorinstabilitybinary, kadokura_2012_rayleightaylorinstabilitytwocomponent}. Recent developments in theoretical \cite{jimbo_2021_surfactantbehaviorthreecomponent} and experimental \cite{eto_2016,bersano_2018} works on three-component BECs have given a new direction to the ongoing research in multi-component condensates. Thus, investigating three-component BECs for their surface and interface properties to explore certain non-linear phenomena, such as interfacial instabilities is aptly aligned with the upsurging research in this domain and fits best for studying the RTI in a controlled environment.
\par
\begin{figure}[t]
    \centering
    \includegraphics[height=0.2\textwidth]{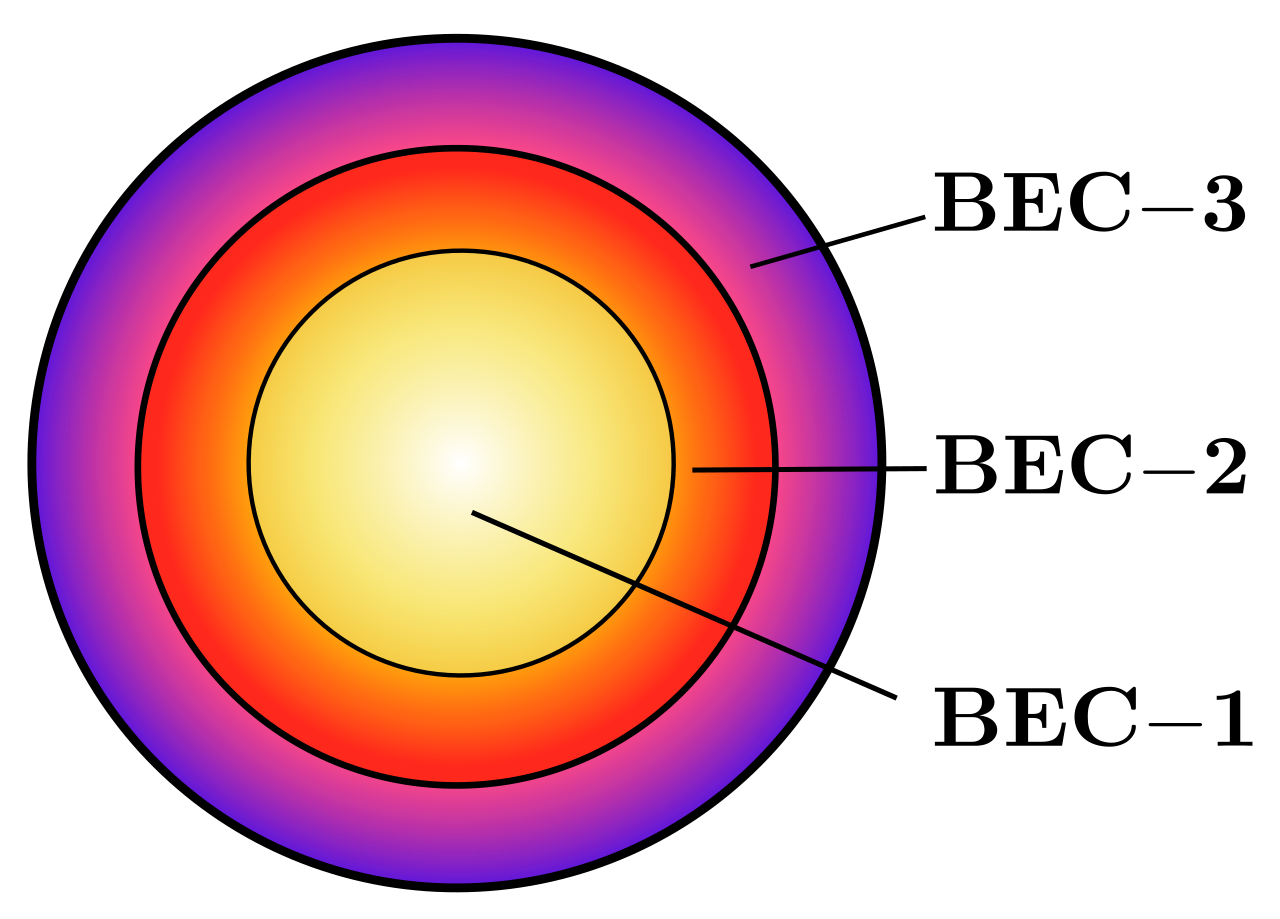}
    \caption{Schematic diagram of a phase separated three-component Bose-Einstein condensate.}
    \label{fig:1}
\end{figure}
In the present paper, we investigate the RTI and subsequent dynamics in a phase-separated three-component BEC confined in an axisymmetric trap. Fig. \ref{fig:1} shows a schematic diagram for the initial arrangement of the components in the system having rotational symmetry. The first component $\mathrm{(BEC{-}1)}$ is depicted by the innermost circle, the second component $\mathrm{(BEC{-}2)}$ surrounds the first component, while the outermost third component $\mathrm{(BEC{-}3)}$ shells the second component. To initiate the RTI, we consider the system to be confined in a quasi-2D harmonic trap. The ground state of the phase-separated three-component BEC with strong repulsive interatomic interactions is chosen as the initial state. Both the interfaces (the first between $\mathrm{BEC{-}1}$ and $\mathrm{BEC{-}2}$, and the second between $\mathrm{BEC{-}2}$ and $\mathrm{BEC{-}3}$) are circular in shape for the axisymmetric trap.  We tune the intraspecies repulsive interaction strengths of one of the three components through Feshbach resonance \cite{roberts_2000_magneticfielddependence} during the dynamical evolution of the system. By doing so, the rotational symmetry of the interface breaks due to the RTI which leads to the deformation of the circular interface into non-linear mushroom fingers. In general, the RTI can be initiated either by modulating the interaction strengths or by altering the trap frequencies \cite{kadokura_2012_rayleightaylorinstabilitytwocomponent}.
Here, we induce the RTI in the following ways: $\rm (i)$ by pushing $\mathrm{BEC{-}3}$ inwards, $\rm (ii)$ by pushing $\mathrm{BEC{-}1}$ outwards, and $\rm (iii)$ by pushing $\mathrm{BEC{-}2}$ outwards.
\par
This paper is organized as follows: Section \ref{sec:2} lays the theoretical foundation and formulation of the problem demonstrating the RTI and ensuing dynamics. We provide the results and discussions in section \ref{sec:3} for three different cases of instigating the RTI, \textit{viz.,} case-\rom{1}, case-\rom{2}, and case-\rom{3} in sections \ref{case:\rom{1}}, \ref{case:\rom{2}}, and \ref{case:\rom{3}}, respectively. We draw important conclusions and provide future perspectives in section \ref{sec:4}. We discuss the normal modes at the interface between two components in Appendix \ref{apx:me}. We give the numerical details used in this paper in Appendix \ref{apx:a}. Appendix  \ref{apx:b} talks about the choice of number of particles in each component. In Appendix  \ref{apx:c}, we highlight a special case of inducing the RTI in the same initial state of the system while we provide a case study on the sudden quench dynamics of the system in Appendix  \ref{apx:d}.
\section{FORMULATION OF THE PROBLEM} \label{sec:2}
We consider a mixture of three kinds of dilute bosonic gases of masses $m_j$ confined in an external axisymmetric potential $V_j$ at zero temperature ($j =1,2,3$). In the mean-field theory, the three-component BEC is described by the governing coupled Gross-Pitaevskii (GP) equations,
\begin{equation} \label{eq:1}
    {\rm i}\hbar\frac{{\partial {\psi_j}}}{\partial t}
    =\bigg[-\frac{\hbar^2}{2 m_j}\nabla^2 + V_j + \sum_{j'=1}^{3}g_{jj'}|\psi_{j'}|^2 \bigg]\psi_j
\end{equation}
where $\psi_j$ is the macroscopic wave function for the $j$-th component, which is normalised as $\int |\psi_j|^2 \dd \mathbf{r} = N_j$ with $N_j$ being the total number of particles in the $j$-th component. The atom-atom interaction is characterised by the parameter $g_{jj'}= 2 \pi \hbar^2 a_{jj'}/m_{jj'}$
where $a_{jj'}$ is the $s$-wave scattering length and $m_{jj'} = m_j m_{j'}/(m_j + m_{j'})$ is the reduced mass between the components $j$ and $j'$ respectively $(j'=1,2,3)$. We ensure that there is no spatial overlap between any two species in accordance with the criteria of phase spearation given as, $g_{jj} g_{j'j'} < g^2_{jj'}$.
\par
We prepare the radially-separated BEC in an axisymmetric harmonic potential, 
\begin{equation} \label{eq:pot}
    V_j= \frac{1}{2}m_j\omega^2(r^2+ \lambda^2z^2)
\end{equation}
where $r$ is the radial distance from the centre of the trap, $\omega$ and $\omega_z$ are the angular trap frequencies in the radial($x-y$ plane) and axial($z$) directions respectively with the trap aspect ratio $\lambda = \omega_z/\omega \gg 1$. We fix the number of atoms in each component $N_j = N $ (see Appendix \ref{apx:b} for details).
\par
The total mean-field energy of the three-component BEC is given as,
\begin{align} \label{eq:8}
    E & = E^K + E^P +\sum_{j} E^I_{jj} + \sum_{j \neq j'}E^I_{jj'} \nonumber \\ & = \int d\mathbf{r} \bigg[\sum_{j'=1}^{3}\bigg(-\psi^{*}_{j}\frac{\hbar^2}{2 m_j}\nabla^2 \psi_{j} + V_j|\psi_{j}|^2 \bigg) \nonumber \\ &+ \frac{1}{2}\sum_{j=1}^{3}g_{jj}|\psi_{j}|^4 + \frac{1}{2}\sum_{j \neq j'}g_{jj'}|\psi_{j'}|^2 |\psi_{j}|^2\bigg]\psi_j
\end{align}

\noindent where the first and second terms respectively contribute to the total kinetic $(E^K)$ and external trapping potential $(E^P)$ energies of the system, the third term contributes to the intraspecies contact interaction energy $E^I_{jj}$ corresponding to the $j${-th} component in the system, and the last term represents the interspecies contact interaction energy $E^I_{jj'}$ contribution owing to any two components ($j \neq j'$) at a certain time.
\par
\paragraph*{Dynamics of three-component BEC in the Thomas-Fermi regime -}
In the Thomas-Fermi (TF) regime, the stationary state density solution of the $j${th} component obeys the relation;
\begin{align} \label{eq:2}
    |\psi_j|^2 = \frac{[\mu_j -V_j]}{g_{jj}}
\end{align}
where the chemical potential of the $j$-th component $\mu_j$ is constrained by the normalisation criteria. The component-wise radial density distributions are given as;
\begin{align} \label{eq:3}
    n_1 = \frac{[\mu_1 -V_1]}{g_{11}},\qquad 0 < r < r_1
\end{align}
\begin{align} \label{eq:4}
    n_2 = \frac{[\mu_2 -V_2]}{g_{22}},\qquad r_1 < r < r_2
\end{align}
\begin{align} \label{eq:5}
    n_3 = \frac{[\mu_3 -V_3]}{g_{33}},\qquad r_2 < r < r_3
\end{align}
where, $r_1$, $r_2$ and $r_3$ represents the radial boundaries of $\mathrm{BEC{-}1}$, $\mathrm{BEC{-}2}$ and $\mathrm{BEC{-}3}$, respectively.
\par
The gradient of the trapping potential, \textit{i.e.,} $\vectorbold{\nabla} V_j$ is analogously treated as the gravitational equivalent for a classical fluid dynamics problem, thereupon the dynamics of the condensate can be modeled as potential flows. Thus, under suitable boundary conditions, a certain combination of the equation of continuity; Euler's equation along with the Bernoulli's theorem can be used to describe the dynamical evolution of the interface between any two fluids in the system \cite{plesset_1954,drazin2004hydrodynamic,chandrasekhar_1981_hydrodynamichydromagneticstability}. A linear stability analysis, for a three-component phase-separated BEC with circular interfaces in the radial plane, manifests that an infinitesimal perturbation at the $p${th} interface (p=1,2) has normal modes of the form $\alpha(s_p) e^{ik_z z + s_p t}$, where $\alpha (s_p)$ is the amplitude of the mode , $k_z$ is the wave number along the $z$ co-ordinate and $s_p$ is the temporal decay constant at the $p$-th interface between any two adjacent components $j, j'$ of densities $n_j$ and $n_{j'}$ respectively. Upon solving the linearized equations for the normal mode \cite{gautam_2010_rayleightaylorinstabilitybinary,plesset_1954,drazin2004hydrodynamic,chandrasekhar_1981_hydrodynamichydromagneticstability,vanschaeybroeck_2009_addenduminterfacetension}, the decay constant $s_p$ can be expressed as;
\begin{align} \label{eq:6}
    s_p = \pm \sqrt{ \Bigg (\frac{k_z \omega^2 r_p (n_{j'} - n_j)}{n_j + n_{j'}} \Bigg)},\qquad j'> j
\end{align}
where $r_p$ demarcates the loci of the $p$-th interface. The stability of any interface largely depends on the \textquoteleft$s_p$\textquoteright \ value associated to it. For imaginary values of $s_p$, (\textit{i.e.,} when $n_j > n_{j'}$), the interface is stable and oscillates about its mean when slightly perturbed. However, for real values of $s_p$ (or when $n_j < n_{j'}$), even infinitesimal perturbations at the interface grow exponentially. Based on this linear stability analysis, we arrive at the sufficient condition required to trigger the RTI in a three-component BEC. Under the TF approximation, for any two components sharing an interface, we have
\begin{align} \label{eq:7}
    a_{j'j'} < a_{jj} \Bigg(\frac{\mu_j' - V}{\mu_{j} - V}\Bigg),\qquad j'> j
\end{align}
for same potential $V$ at the interface. Thus, to set up the RTI in the system of an immiscible mixture of phase-separated three-component BEC, we tune the s-wave scattering lengths through magnetic Feshbach resonance until  the above mentioned criteria is achieved.
\section{NUMERICAL RESULTS} \label{sec:3}

\begin{figure}[H]
    \centering
    \includegraphics[width=0.5\textwidth]{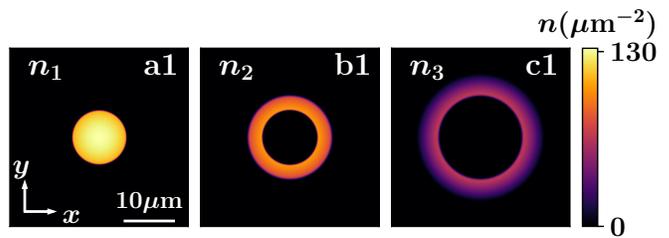}
    \caption{Ground state density profile of a phase-separated three-component Bose-Einstein condensate in an axisymmetric trap with $(\omega, \omega_z) = 2\pi \times (50,2500) \mathrm{Hz}$ and number of atoms in each component, $N_j = 60000$ ($j=1,2,3$).}
    \label{fig:2}
\end{figure}

In this paper, we propose a three-component BEC system of $^{87}\mathrm{Rb}$ - $^{85}\mathrm{Rb}$ - $^{87}\mathrm{Rb}$ mixture confined in a quasi-2D harmonic trap (Eq. \ref{eq:pot}) with trapping frequency $\omega = 2 \pi \times 50$ Hz and $\lambda = 50$. Here, the $\ket {F=2, m_F = -2}$ state of $^{85}\mathrm{Rb}$ ($\mathrm{BEC{-}2}$) is sandwiched between the $\ket {F=1, m_F = -1}$ ($\mathrm{BEC{-}1}$) and $\ket {F=1, m_F = 1}$ ($\mathrm{BEC{-}3}$) hyperfine states of $^{87}\mathrm{Rb}$ as shown in Fig. \ref{fig:2}.
 The experimental realisation of this system can be motivated by similar experiments done for binary BECs \cite{papp_2008_tunablemiscibilitydualspecies,vankempen_2002_interisotopedeterminationultracold,kaufman_2009_radiofrequencydressingmultiple} in which controlled phase-separation has been achieved by altering the intraspecies scattering lengths of either of the components using Feshbach resonance technique. In agreement to these experiments, the intraspecies $s$-wave scattering lengths are chosen to be as; $a_{11}{=}92.4 a_0$, $a_{22}{=}94.5 a_0$ and  $a_{33}{=}100.4 a_0$ with $a_0$ being the Bohr radius.  In our study, we arrange the system such that $a_{11} < a_{22} < a_{33}$. Also, as the mass-difference between any two consecutive components is $\approx 2.3 \%$ 
 \footnote{We prepare this system assuming that the components are concentric in order to compensate for the gravitational sag.},
the arrangement of the components in the system becomes mass independent. Therefore, for a system as shown in Fig. \ref{fig:2}, the ratio $a_{jj}/m_j$ ($j=1,2,3$) should be in increasing order. For the present, it is $1.062, 1.111$ and $1.154$ for $\mathrm{BEC{-}1}$, $\mathrm{BEC{-}2}$  and $\mathrm{BEC{-}3}$ respectively. To prepare the condensates in immiscible regime, the intraspecies scattering lengths are chosen as $a_{12}= a_{21}=213 a_0$, $a_{13} = a_{31}= 213 a_0$, $a_{23}= a_{32}=127 a_0$.
\par
We begin with this state (see Fig. \ref{fig:2}) as the stationary state to initiate the RTI in a multi-component BEC system and examine the dynamical evolution of this system. We first numerically solve the GP Eq. (\ref{eq:1}) (see Appendix \ref{apx:a} for details on numerical implementation)
to generate the ground state of the system.  With this 
{\unskip\parfillskip 0pt \par}
\noindent prepared ground state, we numerically solve the same Eq. (\ref{eq:1}) allowing the system to evolve with time. While examining the time-dynamics for the system, we tune the repulsive intraspecies interaction between components, such that the entire system becomes energetically unfavourable and reaches a metastable state. To induce the RTI to the system, we perturb the phase-separated three-component BEC as shown in Fig. \ref{fig:2}, in three possible ways, viz., (\rm{i}) by decreasing the scattering length $a_{33}$ (case-\rom{1}) in section \ref{case:\rom{1}}, (\rm{ii}) by increasing the scattering length $a_{11}$ (case-\rom{2}) in section \ref{case:\rom{2}}, (\rm{iii}) by increasing the scattering length $a_{22}$ (case-\rom{3}) in section \ref{case:\rom{3}}. We also discuss a special case of decreasing $a_{22}$ as an alternate way to induce RTI in Appendix \ref{apx:c}, and a case of sudden quench dynamics in $\mathrm{BEC{-}2}$ is discussed in Appendix \ref{apx:d}.
\subsection{Case-I: Instability induced by decreasing $a_{33}$} 
\begin{figure}[H]
    \centering
    \includegraphics[scale=1.5,width=\linewidth]{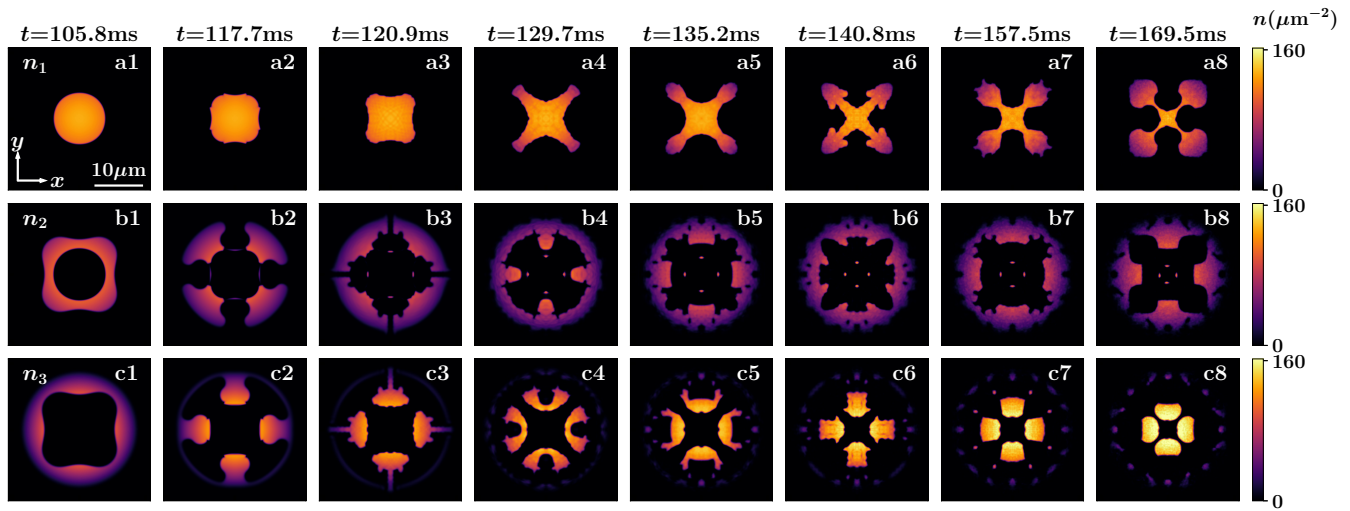}
    \caption{(a1-a8), (b1-b8) and (c1-c8) shows the density profiles $n_1$, $n_2$ and $n_3$ of components $\mathrm{BEC{-}1}$, $\mathrm{BEC{-}2}$ and $\mathrm{BEC{-}3}$ respectively, at different time instances shown in columns: (1) $t = \tau \prime = 105.8 \mathrm{ms} $, (2) $t = \tau = 117.7 \mathrm{ms} $, (3) $t= 120.9 \mathrm{ms} $, (4) $t= 129.7 \mathrm{ms}$, (5) $t=135.2 \mathrm{ms} $, (6) $t=140.8 \mathrm{ms} $, (7) $t=157.5 \mathrm{ms} $, and (8) $t=169.5 \mathrm{ms} $. The scattering length $a_{33}$ is linearly decreased from $100.4 a_0$ to $50 a_0$ between $t=0\mathrm{ms}$ to $t=200 \mathrm{ms}$, after which it is fixed to $50 a_0$. The three-component BEC is initialized in an axisymmetric trap with $(\omega, \omega_z) = 2\pi \times (50,2500) \mathrm{Hz}$ with number of atoms in each component, $N_j = 60000$ ($j=1,2,3$).}
    \label{fig:3}
\end{figure}

\label{case:\rom{1}}
For this case, we decrease the repulsive intraspecies s-wave scattering length $a_{33}$ from $100.4 a_0$ to $50 a_0$ on a linear ramp of ramp time $200 \mathrm{ms}$ and then the system is allowed to evolve further with time. When $a_{33}$ is decreased to $73.7 a_0$, the translational symmetry of the system is broken (at t $\approx$ 105.8 ms), the interface between $\mathrm{BEC{-}2}$ and $\mathrm{BEC{-}3}$ starts to get modulated due to the RTI. However, the interface between $\mathrm{BEC{-}1}$ and $\mathrm{BEC{-}2}$ remains unaffected at this instant (see Fig. \ref{fig:3}, (a1,b1,c1)). As we further decrease $a_{33}$ to 70.7 $a_0$, both the interfaces are affected at about $t=117.7 \mathrm{ms}$ (Fig. \ref{fig:3}, (a2,b2,c3)) and we observe that a 4-fold mushroom-shaped pattern starts to arise in $\mathrm{BEC{-}2}$ (Fig. \ref{fig:3}(b2)) and $\mathrm{BEC{-}3}$ (Fig. \ref{fig:3}(c2)) which grows further with the time propagation, with a perceptible change in the density profile of $\mathrm{BEC{-}1}$. These interface modulations ultimately grows into complicated non-linear 4-fold mushroom like structures called the \textit{Rayleigh-Taylor} (RT) fingers. These fingers arise in all the components of the three-component BEC owing to modulations at the interfaces. As discussed earlier, in the present case, the interface shared between $\mathrm{BEC{-}1}$ and $\mathrm{BEC{-}2}$ is affected much later after the RTI effects are observed at the interface shared between $\mathrm{BEC{-}2}$ and $\mathrm{BEC{-}3}$. This is in accordance with the adiabatic  
\vspace{2 pt}
    {\unskip\parfillskip 0pt \par}

\onecolumngrid

\twocolumngrid
\noindent decrease in the inter-atomic scattering length $a_{33}$ of the outermost component of the ground state shown in Fig. \ref{fig:2}. In Fig \ref{fig:3}, (for later time instances) we show that the density profile of the multi-component BEC is majorly affected due to the translational symmetry breaking as a consequence of RTI. These 4-fold RT fingers corresponds to the normal modes at the interface representing the second order cosine-elliptic Mathieu functions (see Appendix \ref{apx:me} for details).  
\par
It is interesting to note that, the RT fingers (mushroom heads) tend to reach either to the centre or the edge of the system. For $\mathrm{BEC{-}3}$, these fingers grow towards the centre in such a way that this component tends to move to the core of the system. The reason for this change in the density profile is the gradual decrease in the repulsive intraspecies interaction strength $g_{33}$.
Subsequently, the constraint of immiscibility compels the other components to move in the opposite sense which is explained by the expected trend in $\mathrm{BEC{-}1}$ which has mushroom heads towards the periphery of the system; and  for $\mathrm{BEC{-}2}$, a certain change is observed in the direction of the tops of the RT fingers, viz., from towards the edge to oppositely towards the centre of the system. In other words, with the time propagation, the change in the density profiles of all the components (as initiated due to RTI), takes place in such a way that the mushroom tops are in the same direction as the motion of the components, \textit{i.e.,} either towards the centre of the system (for the component moving inwards), or towards the periphery (for the component moving outwards) and that for the intermediate component we observe a sheer change in the direction only after the interface between the $\mathrm{BEC{-}1}$ and $\mathrm{BEC{-}2}$ is modulated due to RTI.
\par
\begin{figure}[t]
    \includegraphics[scale=1.5,width=\linewidth]{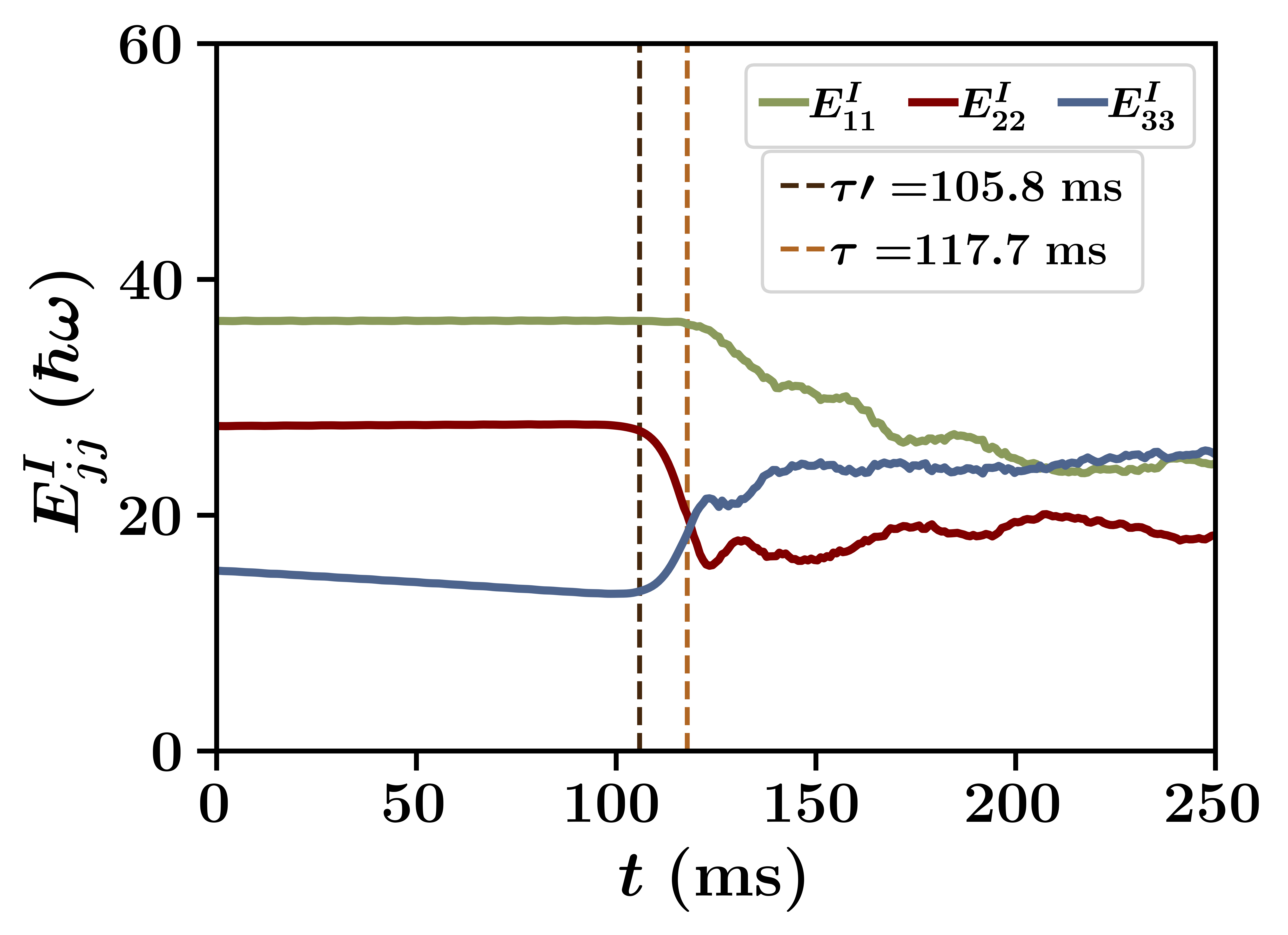}
    \caption{ Variation of contact interaction energies $E^{I}_{11}$(green), $E^{I}_{22}$(red), $E^{I}_{33}$(blue) with the real time propagation of phase separated three-component BEC for the case of \textit{Rayleigh-Taylor} instability seeded by the adiabatic decrease in s-wave intraspecies scattering length $a_{33}$ of the outermost component of the system as shown in Fig. \ref{fig:2}.}
    \label{fig:4}
\end{figure}
This behaviour of the components upon the onset of the seed which induces the RTI in the system, is conformed by the variation of intraspecies contact interaction energies $E^{I}_{jj}$ ($j=1,2,3$) with time as shown in Fig. \ref{fig:4}. It depicts the competition between the various contact interaction energies for the peculiar case under consideration. These contributions are overall constant for no perturbations in the system during the time dynamics. This is in accordance with the stability of the phase-separated three component BEC. However, since we disturb the system by linearly decreasing $a_{33}$, these energy contributions evidently vary with time.
\par
\begin{figure}[t]
    \centering
    \includegraphics[scale=1.5,width=\linewidth]{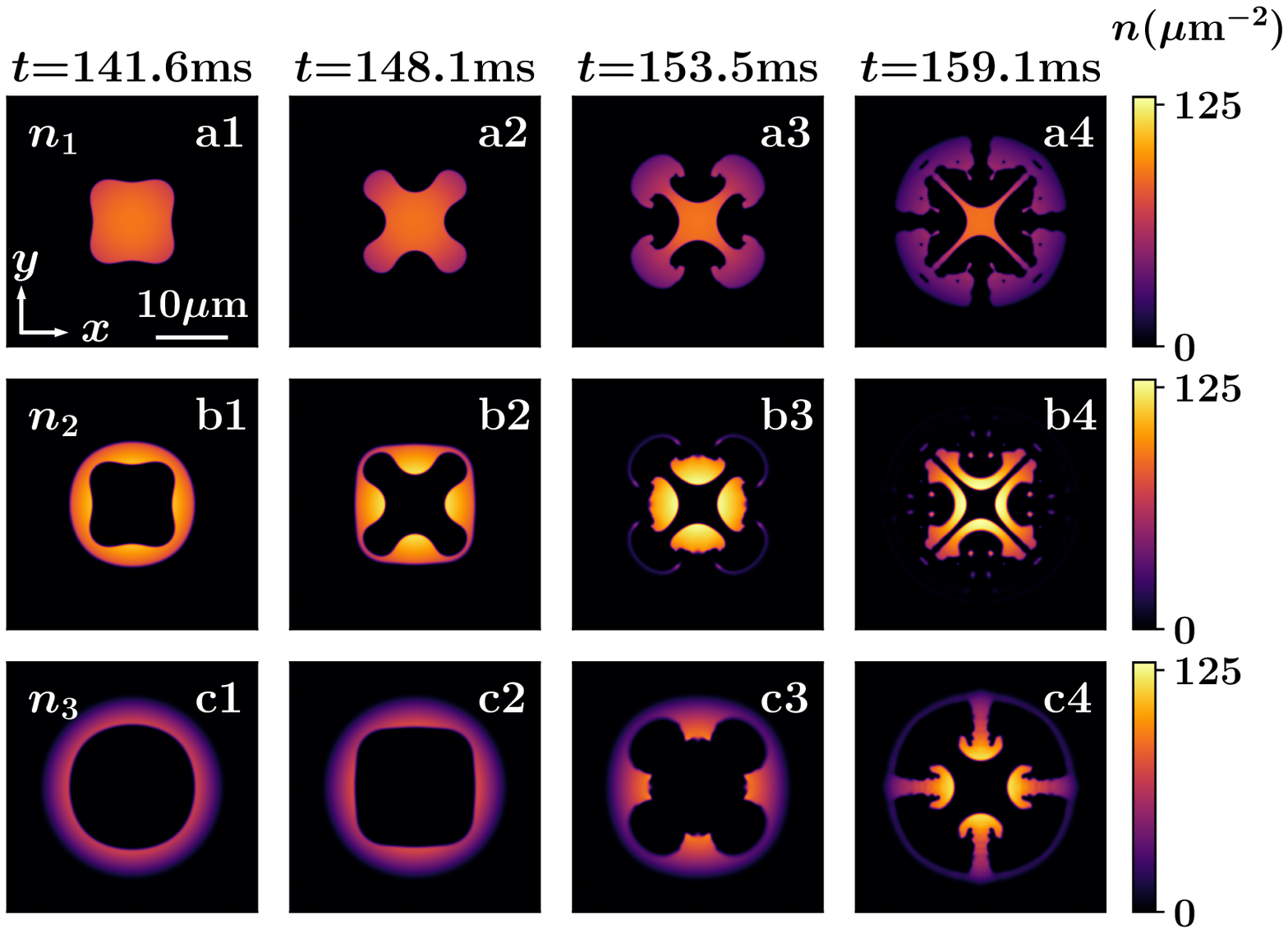}
    \caption{(a1-a4), (b1-b4) and (c1-c4) shows the density profiles $n_1$, $n_2$ and $n_3$ of components $\mathrm{BEC{-}1}$, $\mathrm{BEC{-}2}$ and $\mathrm{BEC{-}3}$ respectively, at different time instances shown in columns: (1) $t =\tau = 141.6 \mathrm{ms} $, (2) $t = 148.1 \mathrm{ms} $, (3) $t = \tau \prime = 153.5 \mathrm{ms} $, (4) $t= 159.1 \mathrm{ms}$. The scattering length $a_{11}$ is linearly increased from $92.4 a_0$ to $240 a_0$ between $t=0\mathrm{ms}$ to $t=200 \mathrm{ms}$, after which it is fixed to $240 a_0$. The three-component BEC is initialized in an axisymmetric trap with $(\omega, \omega_z) = 2\pi \times (50,2500) \mathrm{Hz}$ with number of atoms in each component, $N_j = 60000$ ($j=1,2,3$).}
    \label{fig:5}
\end{figure}
As discussed prior, for the present case it is the second interface (between $\mathrm{BEC{-}2}$ and $\mathrm{BEC{-}3}$) that gets affected the first. We mark this instant as $\tau \prime$ which represents the onset of deformations at the second interface and it is only after this time instant, the density modulations are observed in the system. It is interesting to note that the two contact interaction energies, \textit{i.e.,} $E^{I}_{11}$ and $E^{I}_{22}$ remain constant while the other $E^{I}_{33}$ decreases very slowly until this time instant $\tau \prime$. This is due to the fact that the repulsive intraspecies characteristic interaction strength $g_{33}$ is decreased albeit no modulations at the interface. However, beyond $\tau \prime$, the criteria for the instability at the second interface is achieved, \textit{i.e.,} $a_{33} < a_{22} (\mu_3 - V)/(\mu_2 - V)$ (see Eq. \ref{eq:7}), thereupon the trends in the energies $E^{I}_{22}$ and $E^{I}_{33}$ as well as the density profiles of the corresponding components are modified. $E^{I}_{22}$ starts to decrease while $E^{I}_{33}$ increases owing to the opposite sense of the mushroom tops in either component. As we let the system to evolve further, $\mathrm{BEC{-}3}$ tries to move to the centre by gradually pushing $\mathrm{BEC{-}2}$ to grow outwards. Collaterally, we observe a gradual change in the first interface between $\mathrm{BEC{-}1}$ and $\mathrm{BEC{-}2}$ at certain instant of time $\tau$ beyond which $E^{I}_{11}$ decreases considerably (see Fig. \ref{fig:4}). When we reach a condition such that $a_{33} < a_{11} (\mu_3 - V)/(\mu_1 - V)$ in accordance with Eq. (\ref{eq:7}), we observe a 4-fold non-linear mushroom pattern in all components of the system during the dynamical evolution. Significantly, the interfaces are modulated after certain time instants only as shown in the Fig. \ref{fig:4}, from which we infer that the contact interaction energies $E^{I}_{jj}$ vary with time for such a perturbation as discussed in case-I.
\par
As an alternate way to induce RTI in the system, we suddenly quench $a_{33}$ from $100.4 a_0$ to $50 a_0$ during the evolution of the system. We observe similar results as that for the linear quench case. The time dynamics of the system under sudden quench is given in the supplementary materials.
\begin{figure}[t]
    \centering
    \includegraphics[scale=1.5,width=\linewidth]{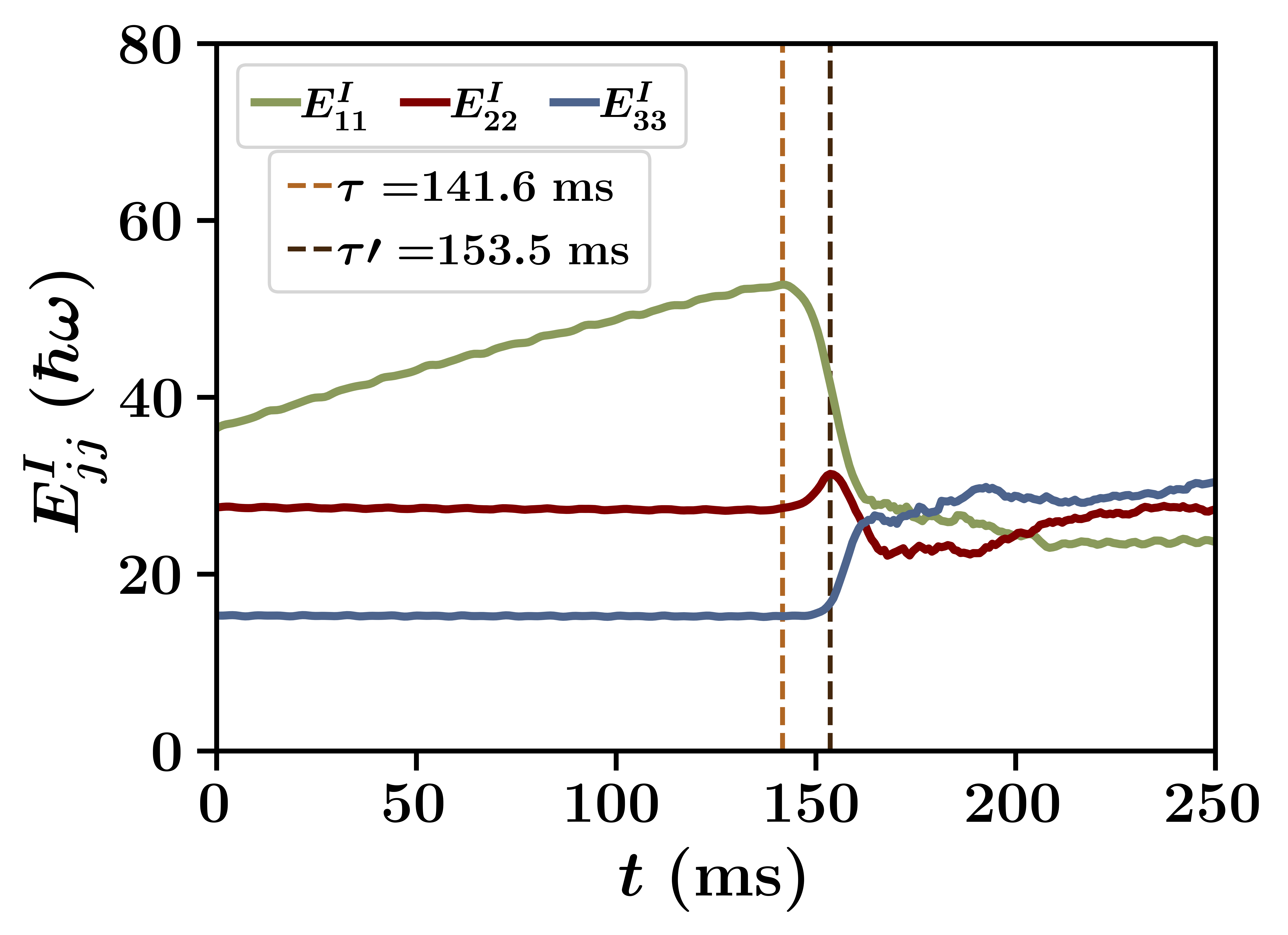}
    \caption{ Variation of contact interaction energies $E^{I}_{11}$(green), $E^{I}_{22}$(red), $E^{I}_{33}$(blue) with the real time propagation of phase separated three-component BEC for the case of \textit{Rayleigh-Taylor} instability seeded by the adiabatic increase in s-wave intraspecies scattering length $a_{11}$ of the outermost component of the system as shown in Fig. \ref{fig:2}.}
    \label{fig:6}
\end{figure}
\subsection{Case-II: Instability induced by increasing $a_{11}$} \label{case:\rom{2}}
In the second case, we increase the repulsive intraspecies characteristic s-wave scattering length $a_{11}$ from $a_{11} = 92.4 a_0$ to $240 a_0$ between the time duration $t= 0 \mathrm{ms}$ to $200 \mathrm{ms}$ and then the system is allowed to dynamically evolve further. The axisymmetry of the system breaks when $a_{11}$ is increased to $196.9 \mathrm{a_0}$ and at this instant, $t = 141.6 \mathrm{ms}$, the first interface between $\mathrm{BEC{-}1}$ and $\mathrm{BEC{-}2}$ starts to get modulated due to the onset of the RTI in the system. However, analogous to {Case-I}, the second interface between $\mathrm{BEC{-}2}$ and $\mathrm{BEC{-}3}$ is yet to note any modulations. And as we further increase $a_{11}$ to $205.7 \mathrm{a_0}$, the second interface between $\mathrm{BEC{-}2}$ and $\mathrm{BEC{-}3}$ gradually starts to get modulated at about $t=153.5 \mathrm{ms}$ beyond which a 4-fold non-linear mushroom-like pattern eventually appears in all the components of the system. Figure \ref{fig:5} shows the density profiles of the three-component BEC at different time instances showing the dynamical evolution of the density profile of either components. These 4-fold patterns denote the normal modes at the interface owing to the second order cosine-elliptical Mathieu function (see Appendix \ref{apx:me}). In this case also, the mushroom tops move either towards the centre or to the edge of the system so as to maintain the constraint of immiscibility throughout the time propagation of the system under consideration. They are  outwards for $\mathrm{BEC{-}1}$ and inwards for both $\mathrm{BEC{-}2}$ and $\mathrm{BEC{-}3}$. As we can see from Fig. \ref{fig:5}, the mushroom-like structures or \textit{Rayleigh-Taylor} fingers grows into definite shape as the RTI sets in. Thence, we highlight that the condensate $\mathrm{BEC{-}1}$ eventually grows outwards and its position is being filled by the condensate $\mathrm{BEC{-}2}$ while the third condensate $\mathrm{BEC{-}3}$ is shelling the other two.
\par

\begin{figure}[t]
    \centering
    \includegraphics[scale=1.5,width=\linewidth]{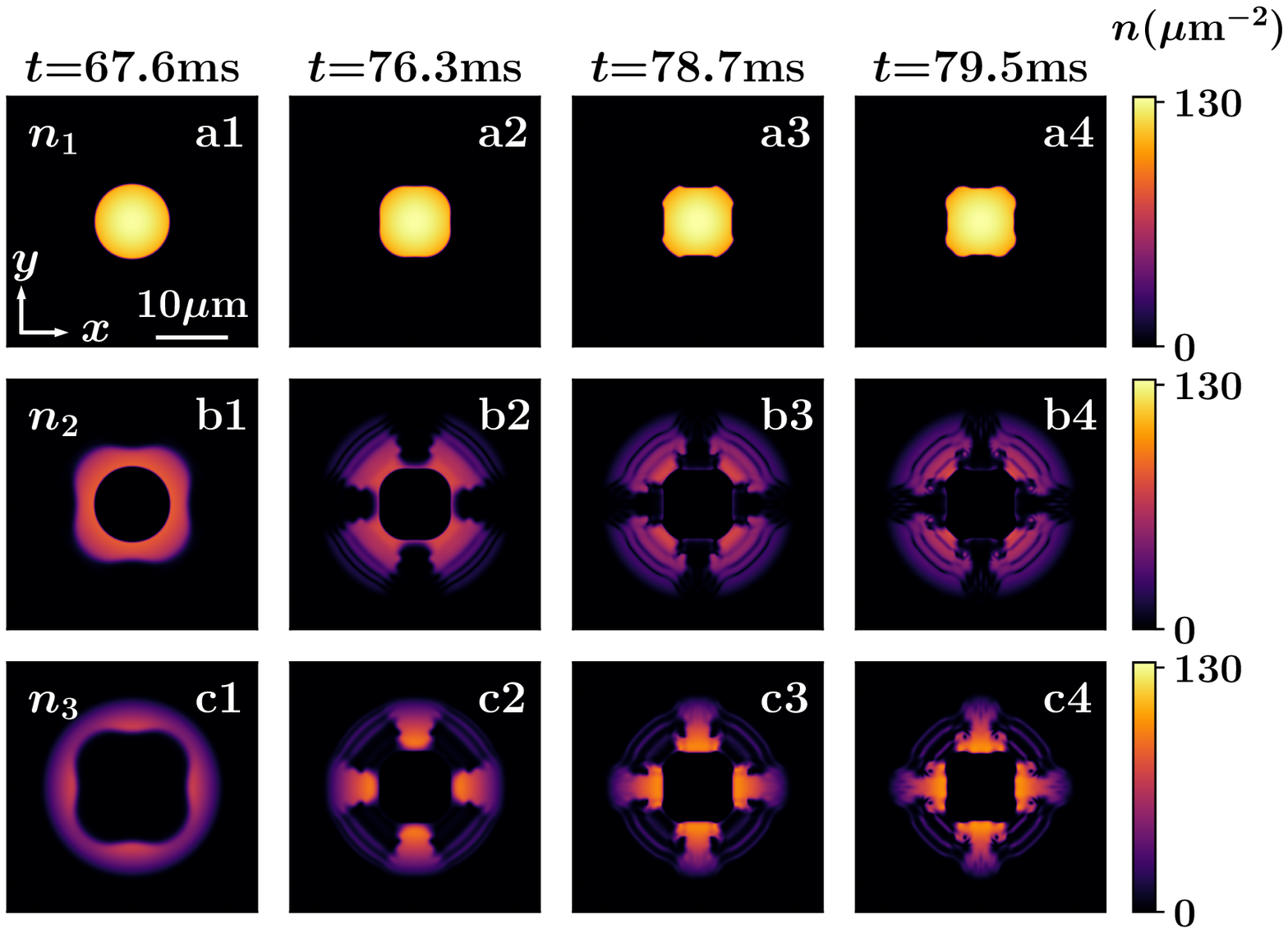}
    \caption{(a1-a4), (b1-b4) and (c1-c4) shows the density profiles $n_1$, $n_2$ and $n_3$ of components $\mathrm{BEC{-}1}$, $\mathrm{BEC{-}2}$ and $\mathrm{BEC{-}3}$ respectively, at different time instances shown in columns: (1) $t = \tau \prime = 67.6 \mathrm{ms} $, (2) $t = 76.3 \mathrm{ms} $, (3) $t = \tau = 78.7 \mathrm{ms} $, (4) $t= 79.5 \mathrm{ms}$. The scattering length $a_{22}$ is linearly increased from $94.5 a_0$ to $240 a_0$ between $t=0$ ms to $t=200 \mathrm{ms}$, after which it is fixed to $240 a_0$. The three-component BEC is initialized in an axisymmetric trap with $(\omega, \omega_z) = 2\pi \times (50,2500) \mathrm{Hz}$ with number of atoms in each component, $N_j = 60000$ ($j=1,2,3$).}
    \label{fig:7}
\end{figure}
We show the variation of contact interaction energies $E^{I}_{11}$, $E^{I}_{22}$ and $E^{I}_{33}$ with time respectively for the present case in Fig. \ref{fig:6}. Here, we mark the onset of RTI videlicet the density modulation at the first interface separating the condensates $\mathrm{BEC{-}1}$ and $\mathrm{BEC{-}2}$ by $\tau$ and that at the second interface separating $\mathrm{BEC{-}2}$ and $\mathrm{BEC{-}3}$ by $\tau \prime$ respectively. These instabilities in the system occur with respect to Eq. (\ref{eq:7}) in correspondance with the modulations across the two interfaces.  From the figure, we state that until time $\tau$, \textit{i.e.,} before RTI is being induced in the system, the interaction energies $E^{I}_{22}$ and $E^{I}_{33}$ are almost constant while $E^{I}_{11}$ increases uniformly as we increase $a_{11}$ linearly. After $\tau$, when the density modulations begin, $E^{I}_{11}$ starts to decrease while $E^{I}_{22}$ gradually increases. This trend in the variation of contact interaction energies goes hand in hand with the density modulations about the first interface. Note that $E^{I}_{33}$ only changes after the density modulations occur about the second interface, \textit{i.e.,} only after $\tau \prime$ following which $E^{I}_{22}$ decreases while $E^{I}_{33}$ increases with time. This behaviour is justified as per the density modulations into 4-fold RT fingers in the system. We observe that the interaction energies saturates to a certain value in the long run which represents a highly unstable stage of the system with the \textit{Rayleigh-Taylor} instability.
\par
We also observe similar effects of RTI in the system when $a_{11}$ is suddenly quenched from $92.4 a_0$ to $240 a_0$. The sudden quench dynamics are given in the supplementary materials.
\begin{figure}[t]
    \centering
    \includegraphics[scale=1.5,width=\linewidth]{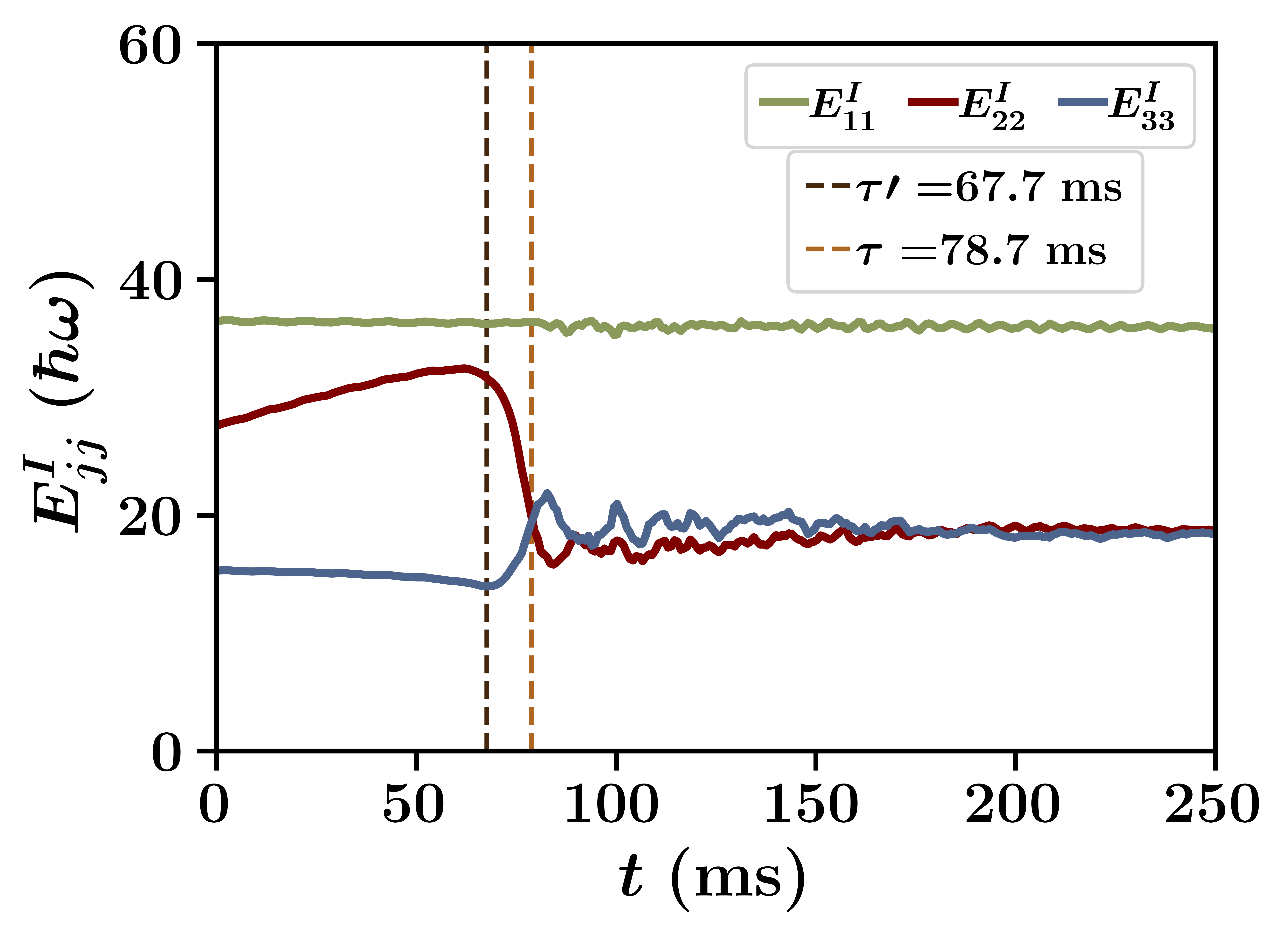}
    \caption{ Variation of contact interaction energies $E^{I}_{11}$(green), $E^{I}_{22}$(red), $E^{I}_{33}$(blue) with the real time propagation of phase separated three-component BEC for the case of \textit{Rayleigh-Taylor} instability seeded by the adiabatic increase in s-wave intraspecies scattering length $a_{22}$ of the outermost component of the system as shown in Fig. [\ref{fig:2}].}
    \label{fig:8}
\end{figure}
\par
\subsection{Case-III: Instability induced by increasing $a_{22}$} \label{case:\rom{3}}
In the third case, we linearly increase the intraspecies s-wave scattering length $a_{22}$ with time in order to induce the \textit{Rayleigh-Taylor} instability in the system. During the dynamical evolution of the ground state as shown in Fig. \ref{fig:2}, we increase $a_{22}$ from $94.5 a_0$ to $240 a_0$ corresponding to $t = 0 \mathrm{ms}$ and $t=200 \mathrm{ms} $ respectively and the system is allowed to dynamical evolve further. Similar to the previous cases, we observe that the translational symmetry of the system breaks at $\tau \prime =67.6 \mathrm{ms}$ when $a_{22}$ is increased to $143.7 a_0$ and the second interface is affected first adhering to which the densities of $\mathrm{BEC{-}2}$ and $\mathrm{BEC{-}3}$
are changed. We recall that the second interface gets dynamically unstable according to the Eq. (\ref{eq:7}) as beyond this time instant $\tau \prime$, $a_{22} > a_{33} (\mu_2 - V)/(\mu_3 - V)$.  As we further increase $a_{22}$ to $151.8 a_0$, the first interface between $\mathrm{BEC{-}1}$ and $\mathrm{BEC{-}2}$ starts to deform at time instant $\tau = 78.7 \mathrm{ms}$. However, as we have chosen the initial state to have $a_{11} < a_{22} < a_{33}$, it is trivial that the condition for instability $a_{22} < a_{11} (\mu_2 - V)/(\mu_1 - V)$ at the first interface between $\mathrm{BEC{-}1}$ and $\mathrm{BEC{-}2}$ will not be achieved which is why the density modulations at the first interface are merely collateral impacts of the second interface modulations due to the RTI. The 4-fold RT fingers begin to arise at all the components of the system much later than these critical time instances $\tau$ and $\tau \prime$.  Figure \ref{fig:7} shows the density profiles at various time instants during the time dynamics of the system as suggested in the present case. In correspondance with the figure, we comform the formation of mushroom-like patterns in the system after it has been triggered. Analogous to the other two cases, the 4-fold non-linear pattern denotes the normal modes for the cosine-elliptic Mathieu functions for $m=2$ as discussed in Appendix \ref{apx:me}. Also, as we linearly increase $a_{22}$, to sustain the immiscibility criteria, the condensate densities move in the following way. The component $\mathrm{BEC{-}2}$ over time comes towards the periphery of the system with $\mathrm{BEC{-}3}$ filling up its void. However, $\mathrm{BEC{-}1}$ remains shelled by the other components throughout the dynamical evolution. Thus, its density is not affected much relative to the other two components. The 4-fold non-linear pattern appears only when the density modulations of $\mathrm{BEC{-}2}$ and $\mathrm{BEC{-}3}$ have already happened. In the long run, the system enters into breathing modes.
\par

In Fig. \ref{fig:8}, we interpret the time variation of the contact interaction energies $E^{I}_{jj}$, $(j=1,2,3)$ for the present case. As we have already discussed, the RTI triggers into the system only after the system is allowed to evolve for a certain time. We associate these critical time instances, beyond which the first and second interfaces begin to get modulated, as $\tau$ and $\tau \prime$ respectively. Here, we analyse the behaviour of energies $E^{I}_{jj}$ in three different time zones, \textit{i.e.,} \paragraph{} \hskip -10pt between $t=0\ \mathrm{ms} $ to $t = \tau \prime\ \mathrm{ms}$ - for which $E^{I}_{11}$ and $E^{I}_{33}$ are constant while $E^{I}_{22}$ increases linearly so as to adjust the increasing repulsive intraspecies intraction $g_{22}$ albeit the density modulation;  \paragraph{} \hskip -10pt between $t= \tau \prime\ \mathrm{ms} $ to $t = \tau\ \mathrm{ms}$ - which marks the induction of RTI in the system as density modulations in $\mathrm{BEC{-}2}$ and $\mathrm{BEC{-}3}$ are observed owing to which we notice that $E^{I}_{22}$ decreases smoothly while $E^{I}_{33}$ gradually increases with a very inconsequential variation in $E^{I}_{11}$ as it is negligibly affected by the perturbation triggering RTI in the system, and \paragraph{} \hskip -10pt beyond $\tau$ - after which the system gradually enters into the breathing modes which is reflected by the saturation of the contact interaction energies $E^{I}_{jj}$ after little fluctuations over time.
\par

\vspace{-0.3cm}
\section{Conclusion} \label{sec:4}
\vspace{-0.2cm}
In conclusion, we have thoroughly examined the various possibilities for the onset of the \textit{Rayleigh-Taylor} instability in a phase-separated three-component BEC. We have proposed a BEC mixture of $^{87}\mathrm{Rb}$ - $^{85}\mathrm{Rb}$ - $^{87}\mathrm{Rb}$ in a quasi-2D axisymmetric trap as a case study and have suggestted certain ways to initiate the RTI in the system.  We numerically solve the GP equations to generate the ground state of the phase-separated three-component BEC. We have discussed three different cases for initiating the RTI in the system. During the dynamical evolution of the system, we have tuned the intra-species $s$-wave scattering lengths of either of the three components. As a result, the stable configuration of the system changes when the intra-species scattering length reaches to a critical value. Subsequently, to restore the new stable configuration, the circular symmetry of the interface breaks. This leads to the growth of mushroom-shaped structures that corresponds to the cosine-elliptic angular Mathieu functions representing the normal modes at the interface. A comparison between all of the above mentioned cases has been presented by drawing similarities in the density modulations. The time variation of the interaction energies rightly compliments the results obtained during the density modulations. The results obtained in either cases are in excellent agreement with one another. Our study provides insights into the dynamics of three-component BECs that have potential implications for the experimental observation of the \textit{Rayleigh-Taylor} instability in multi-component BECs.
\par
Our findings pave the way for several future research avenues in multi-component BEC. In this work, we have restricted our studies to a particular case where each condensate has the same number of particles. However, it would be intriguing to explore the effect of different ratios of the number of particles on the interfacial dynamics. Furthermore, most studies on \textit{Rayleigh-Taylor} instability have been done in 2D. So there is a growing interest in exploring the phenomenon in three dimensions. This could provide a better understanding to the full complexity of the instability and its behavior in different geometries. Moreover, magnetic fields can be used to control the behavior of atomic BECs in different hyperfine states \cite{eto_2016}. Incorporating magnetic fields into the study of \textit{Rayleigh-Taylor} instability in multi-component BECs could provide new avenues for controlling and manipulating the instability. Also, multi-component BEC qualify themselves to be a favoured candidate to study various interfacial and other non-linear phenomena in quantum fluids, thereby renewing interests for researchers to look into challenging problems mimicking other kinds of instabilities such as \textit{Kelvin-Helmholtz} instability \cite{blaauwgeers_2002_shearflowkelvinhelmholtz,takeuchi_2010_quantumkelvinhelmholtzinstability,suzuki_2010_crossoverkelvinhelmholtzcountersuperflow}, \textit{Richtmyer-Meshkov} instability \cite{jourdan_2005,zhou_2019}, \textit{Plateau-Rayleigh} instability in the future.
\section*{Acknowledgements}
We acknowledge the PARAM Shakti
(Indian Institute of Technology Kharagpur)—a national supercomputing mission, Government of India for providing
computational resources. A.S. gratefully acknowledges the support from the Prime Minister's Research Fellowship (PMRF), India. S.H. acknowledges the MHRD, Govt. of India for the research fellowship. We acknowledge Dr. Koushik Mukherjee for useful discussions.

\appendix

\section{Normal modes} \label{apx:me}
The system considered here is a zero-temperature phase-separated three-component BEC. The stationary state of the system can be well described by the time-independent GP equations,
\begin{equation}
	\left[-\frac{\hbar^2}{2m_j}\nabla^2+V_j+\sum_{j^{\prime}=1}^{3}g_{jj^{\prime}}\abs{\psi_{j^{\prime}}}^2\right]\psi_j=\mu_j \psi_j.\label{GPE}
\end{equation}
As we discussed in sec \ref{sec:2} of the main text, we consider a quasi-two-dimensional harmonic potential of the form 
\begin{equation}
	V_{j}=\frac{1}{2}m_j\omega^2(r^2+\lambda^2 z^2),
\end{equation}
where $r^2 = x^2+y^2$ and the trap aspect ratio $\lambda=\omega_z/\omega>>1$. Since, the density of each species are relatively low at the interface, we can neglect the intra and inter-species interactions at the interface. Under such circumstances and in the quasi-2D limit the Eq. \ref{GPE} takes the form,
\begin{align}
	&\left(-\nabla_{\perp}^2+V_j\right)\psi_j={\mu_j}\psi_j\nonumber\\
	\implies&\left(\nabla_{\perp}^2+K_j^2\right)\psi_j=0\label{Helm}.
\end{align} 
where $K_j^2 = {\mu_j} - V_j $.
\par
The Eq. \ref{Helm} is the 2D-Helmoltz equation. Now if we do a co-ordinate transformation from 2D cartesian coordinates $(x,y)$ to elliptic coordinates $(r,\theta)$ defined by the relations $x=a\cosh r\cos\theta$, $y=a\sinh r\sin\theta$ where $r>0$ and $\theta\in(0,2\pi)$, the Eq. \ref{Helm} can be cast into the following form; 
\begin{equation}
	\frac{1}{a^2(\sinh^2r+\sin^2\theta)}\left(\pdv[2]{\psi_j}{r}+\pdv[2]{\psi_j}{\theta}\right)+k_j^2\psi_j=0\label{helm2}
\end{equation}
with the solution $\psi_j$ given as $\psi_j=R(r)\Theta(\theta)$. Now substituting $\psi_j$ in Eq. \ref{helm2} we get,
\begin{equation}
	\left(\frac{1}{R}\pdv[2]{R}{r}+k_j^2a^2\sinh^2r\right)+\left(\frac{1}{\Phi}\pdv[2]{\Phi}{\theta}+k_j^2a^2\sin^2\theta\right)=0\label{helm3}
\end{equation}
Now by the method of separation of variables, Eq. \ref{helm3} can be reduced to the radial and the angular Mathieu equations \cite{roy_2012, gutierrez2003mathieu} viz.,

\begin{align}
	\dv[2]{R}{r}-\left[\left(A+\frac{k_j^2a^2}{2}\right)-\frac{k_j^2a^2}{2}\cosh2 r \right]R&=0, \label{RME}\\
	\dv[2]{\Theta}{\theta}+\left[\left(A+\frac{k_j^2a^2}{2}\right)-\frac{k_j^2a^2}{2}\cos2\theta\right]\Theta&=0 \label{AE}
\end{align} 
where $A$ is the separation constant. The interface of the phase-separated three-component BEC has fixed radial coordinate $r$. However, the angle $\theta$ varies from 0 to $2\pi$. Thus the solutions, $\Theta$ of the angular Mathieu equation (Eq. \ref{AE}) represent the normal modes. Now for a circular symmetric trapping potential, the angular (Eq. \ref{AE}) and the radial Mathieu equations (Eq. \ref{RME}) becomes the well known harmonic and Bessel equation in the limit $a\to0$, respectively. In this limit Eq. \ref{AE} becomes
\begin{equation}
	\dv[2]{\Theta}{\theta}+A\Theta=0\label{har},
\end{equation} 
and the solutions of Eq. \ref{har} are the trigonometric function $\cos (m\theta)$ (even) and $\sin (m\theta)$ (odd), where $m$ is the order of the angular Mathieu function.
\par
As discussed in section \ref{sec:3} for case-\rom{1}, case-\rom{2} and case-\rom{3} in sections \ref{case:\rom{1}}, \ref{case:\rom{2}} and \ref{case:\rom{3}} respectively, the 4-fold symmetric mushroom-shaped pattern develops as the \textit{Rayleigh-Taylor} instability sets in. The corresponding solution of Eq. \ref{har} which satisfies this 4-fold symmetric pattern is $\cos2\theta$. We discuss in  Appendix \ref{apx:c} a case of instability corresponding to the cosine-elliptic Mathieu function od order 1 for which a 2-fold mushroom-shaped pattern is observed. These normal modes grow exponentially with time according to the Eq. {\ref{eq:7}}.

\section{Details on numerical implementation} \label{apx:a}
In section \ref{sec:3} of the main text, we mentioned about numerically solving Eq. (\ref{eq:1}) to generate the ground state and thereby examining the dynamical evolution of its solution under certain perturbations so as to study the RTI in the phase-separated three-component BEC. In our case study, we consider the system to be confined in a quasi-2D harmonic trapping potential. In the quasi-2D regime, the motion of atoms in the $z$-direction are rendered insensitive and the wave functions $\psi_j$ ($j=1,2,3$) are expressed as $\phi_j(x,y) \zeta(z)$ where $\zeta(z) = (\lambda / \pi)^{1/4} \mathrm{exp}(-\lambda z^2 /2)$ is the ground state solution along the $z$-direction and $\lambda$ is the trap aspect ratio. After integrating out the variable $z$, the 2D dimensionless time-dependent GP Eq. (\ref{eq:1}) governing the dynamics of the BEC takes the form \cite{pethick_smith_2008, pitaevskii_bose-einstein_2003};
\begin{align} \label{eq:DE}
    {\rm i}\frac{{\partial {\phi_j}}}{\partial t}
    =\bigg[{-}\frac{1}{2}\frac{m_0}{m_j}\nabla_{\perp}^2 + \frac{1}{2}\frac{m_j}{m_0}(x^2 + y^2) + \sum_{j'=1}^{3}\mathcal{G}_{jj'}N_j|\phi_{j'}|^2 \bigg]\phi_j
\end{align}
where $m_0$ is the minimum of atomic masses {$m_j$} and $\mathcal{G}_{jj'} = \sqrt{\lambda/2\pi}2\pi a_{jj'} m_0/m_{jj'}$ is the effective 2D nonlinear interaction. We prepare the initial state by rescaling the lengths in the unit of the characteristic length given as $a_{osc} = \sqrt{\hbar / {m_0 \omega}}$ and $\omega$ is the radial trapping frequency. To generate the ground state, we numerically solve the Eq. (\ref{eq:DE}) using the split-step Crank-Nicholson method by propagating the macroscopic wave functions $\psi_j$ in imaginary time \cite{crank_nicolson_1947,ANTOINE20132621,muruganandam_2009_fortranprogramstimedependent}. Next, we use this imaginary timme propagated ground state as our initial state and solve the same Eq. (\ref{eq:DE}) allowing the system to evolve in real time for a specific case of perturbation as discussed in the main text. All of our simultions runs from a spatial extent of $-20 a_{osc}$ to $20 a_{osc}$ in both $x$ and $y$ (radial) directions with $2001 \times 2001$ grid points. The employed spatial discretization (grid spacing) refers to $\Delta x = \Delta y = 0.02\ a_{osc}$, while the time step of the numerical integration is $\delta t = (2 \times 10^{-4})/{\omega}$.

\section{ Discussion on number of particles in each component} \label{apx:b}

In the main text, we chose the number of particles in the ratio 1:1:1 in each component of the three-component BEC to prepare the ground state. 
%
{\unskip\parfillskip 0pt \par}
\begin{figure}[t]
    \centering
    \includegraphics[scale=1.5,width=\linewidth]{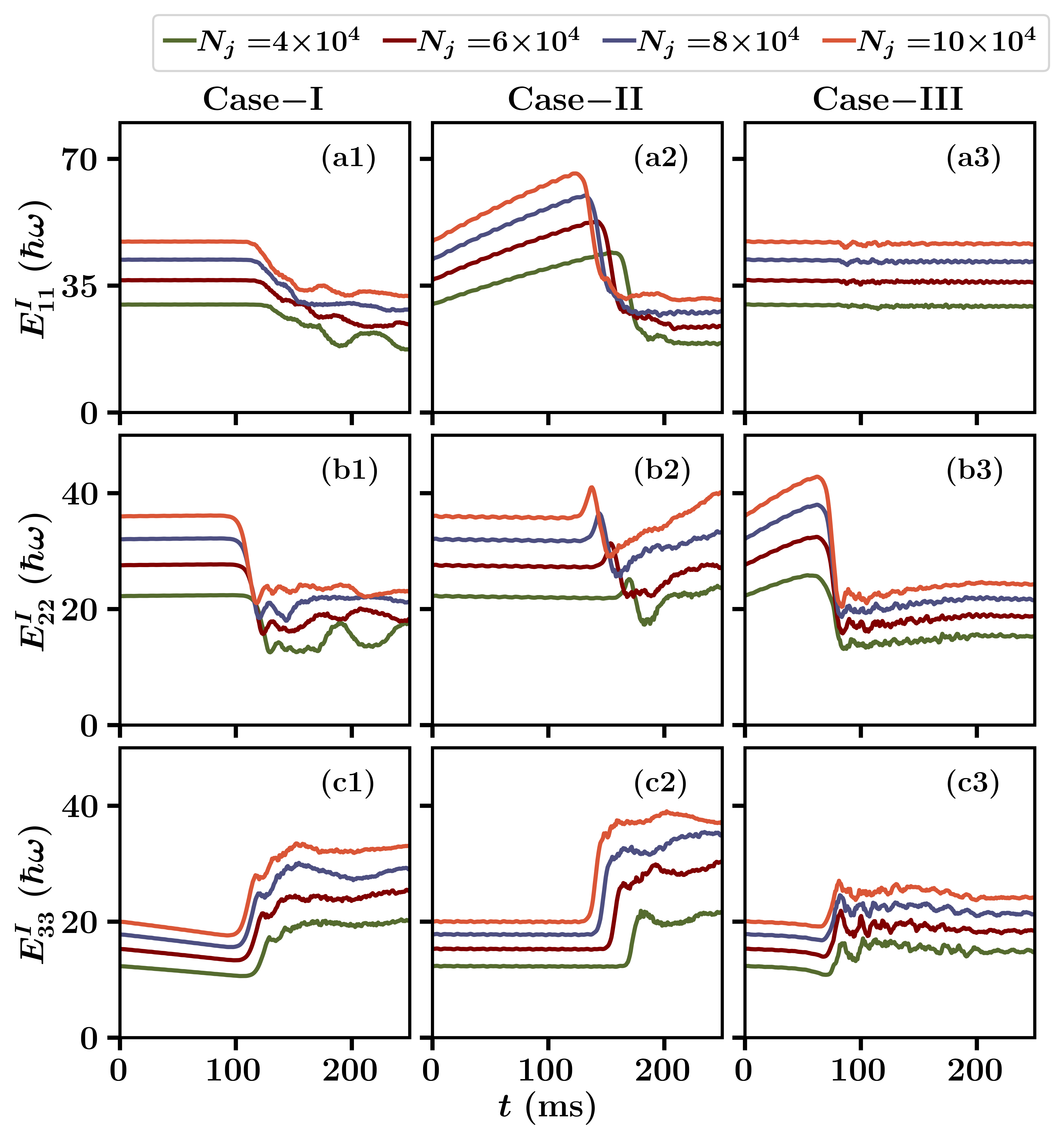}
    \caption{ Variation of contact interaction energies $E^{I}_{jj}$ (rows [1-3]) with the time propagation of phase separated three-component BEC for all the cases (columns [1-3]) of \textit{Rayleigh-Taylor} instability discussed in the main text for different number of particles $N_j$ in the system.}
    \label{fig:9}
\end{figure}
\noindent In this section, we draw a comparison between the various cases for triggering the RTI in the system, discussed above for different number of particles in the same ratio. Much likely in the same way as discussed in the text, we generate the initial ground state of the system with different values of total number of particles per component, viz., for $N_j = (4,6,8,10)\times 10^4$, by numerically solving the Eq. (\ref{eq:1}). We further study the time propagation of each of these initial states for all the three cases mentioned for inducing the RTI in the system.
\par
{\unskip\parfillskip 0pt \par}
\begin{figure}[tb!]
    \centering
    \includegraphics[scale=1.5,width=\linewidth]{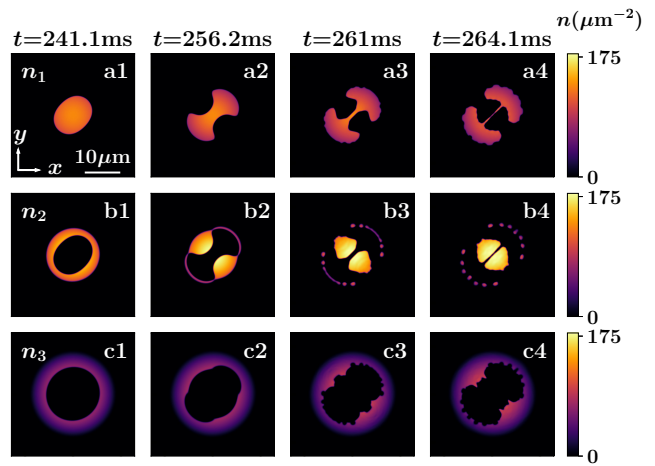}
    \caption{(a1-a4), (b1-b4) and (c1-c4) shows the density profiles $n_1$, $n_2$ and $n_3$ of components $\mathrm{BEC{-}1}$, $\mathrm{BEC{-}2}$ and $\mathrm{BEC{-}3}$ respectively, at different time instances shown in columns: (1) $t = \tau = 241.1 \mathrm{ms} $, (2) $t = \tau \prime= 256.2 \mathrm{ms} $, (3) $t = 261 \mathrm{ms} $, (4) $t= 264.1 \mathrm{ms}$. The scattering length $a_{22}$ is linearly decreased from $94.5 a_0$ to $50 a_0$ between $t=0\mathrm{ms}$ to $t=100 \mathrm{ms}$, after which it is fixed to $50 a_0$. The three-component BEC is initialized in an axisymmetric trap with $(\omega, \omega_z) = 2\pi \times (50,2500) \mathrm{Hz}$ with number of atoms in each component, $N_j = 60000$ ($j=1,2,3$).}
    \label{fig:10}
\end{figure}
We compare all of these cases in Fig. \ref{fig:9} in which we show the variation of ${E^I_{jj}}$ with $N_j$ for either case. From the figure, we infer that each of $E^I_{jj}$ follows a similar trend when varied with time under either case of perturbation which strongly agrees with the choice of the number ratio. In Fig. \ref{fig:9}, each vertical column represents the three cases of inducing the RTI in the system as discussed in the main text and in each row, we compare the $E^{jj}$'s for different number of particles. Here, we observe that for a particular case of perturbation, the time-variation of the contact interaction energies is typical and the trend is exactly same for different number of particles. In other words, this choice of the initial state to work with for triggering instability in the system is independent of the number of particles $N_j$. Thence, we can comply that for any system as shown in [Fig. \ref{fig:2}], we can induce the RTI by tuning the scattering lengths independently.

\section{ \textit{Rayleigh-Taylor} instability following a linear quench in \textbf{${a_{22}}$}} \label{apx:c}
In the main text, we discussed the conventional methods for inducing the RTI in a phase-separated three-component BEC. We studied the dynamical evolution of the system under the three cases in great details. In this section, we focus on a special case to trigger the instability in the system. Here, we decrease the s-wave scattering length $a_{22}$ from ${94.5\ a_0}$ to ${50\ a_0}$ for a period of 100 ms and then the system is allowed to propagate in time.

For such a perturbation, we observe the density modulations appearing at the interfaces which grows exponentially into complicated non-linear fingers much like the previously
 discussed cases. The only contrast being the much delayed episode of instability that too with the occurance of a 2-fold mushroom instability at the interfaces instead of a 4-fold pattern. We have shown the snapshots of the density profiles of the three-component BEC in Fig. \ref{fig:10}.  From this figure, we infer that the onset of density modulations first start at  the interface shared between $\mathrm{BEC{-}1}$ and $\mathrm{BEC{-}2}$ at around $t= 241.1 \mathrm{ms}$, and later at around $t= 256.2 \mathrm{ms}$, the interface between $\mathrm{BEC{-}2}$ and $\mathrm{BEC{-}3}$ starts to get deformed. These deformations grow into non-linear fingers in the long run as shown in Fig. \ref{fig:10}.
\par

It is both interesting and important to appreciate the subtle differences between this case and those discussed in the main text, particularly the third case. In this case, as we gradually decrease the s-wave scattering length $a_{22}$ thereby decreasing the repulsive intraspecies interaction $g_{22}$, the condensate $\mathrm{BEC{-}2}$ over time tries to come to the centre of the system. However, since $\mathrm{BEC{-}3}$ is shielding the other two components, it is minimally affected due to the RTI induced in this case for the same reasons due to which the condensate $\mathrm{BEC{-}1}$ remains relatively less affected in \textit{Case-\ref{case:\rom{3}}}. Another noteworthy feature of this case is the contact interaction energies $E^{I}_{11}$ and $E^I_{33}$ are almost constant until the onset of the RTI in the system, while $E^I_{22}$ first decreases upto 100 ms (during which the scattering length $a_{22}$ is being decreased linearly), and after that it is almost constant until the density modulations begin. We have not shown the figure for contact energy variations here.
\section{Instability following a sudden quench of \textbf{${a_{22}}$}} \label{apx:d}
For this case, we induce the instability in a phase-separated system (Fig. \ref{fig:2}) with $N_j=60000$, chosen as 
\begin{figure}[H]
    \includegraphics[scale=1.5,width=\linewidth]{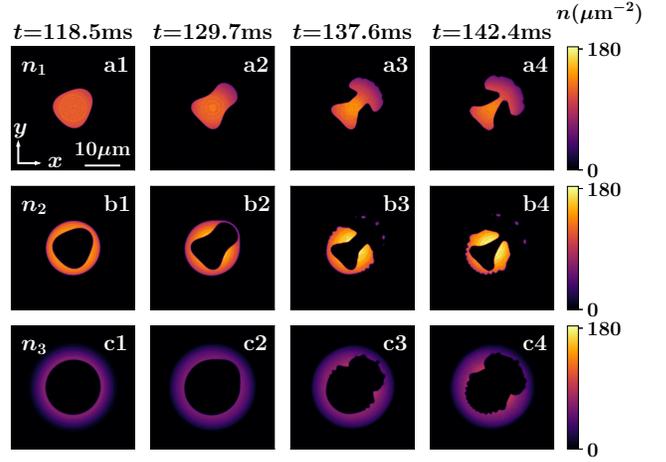}
    \caption{(a1-a4), (b1-b4) and (c1-c4) shows the density profiles $n_1$, $n_2$ and $n_3$ of components $\mathrm{BEC{-}1}$, $\mathrm{BEC{-}2}$ and $\mathrm{BEC{-}3}$ respectively, at different time instances shown in columns: (1) $t = \tau = 118.5 \mathrm{ms} $, (2) $t = \mathcal{\tau \prime} = 129.7 \mathrm{ms} $, (3) $t  = 137.6 \mathrm{ms} $, (4) $t= 142.4 \mathrm{ms}$. The scattering length $a_{22}$ is suddenly decreased from $92.4 a_0$ to $50 a_0$, after which it is fixed to $50 a_0$. The three-component BEC is initialized in an axisymmetric trap with $(\omega, \omega_z) = 2\pi \times (50,2500) \mathrm{Hz}$ with number of atoms in each component, $N_j = 60000$ ($j=1,2,3$).}
    \label{fig:11}
\end{figure}
\noindent
the initial state. Here, we suddenly quench the $s$-wave scattering length $a_{22}$ from ${94.5 a_0}$ to ${50 a_0}$ during the time propgation and let the system to dynamically evolve further in a similar way as we discussed for the other cases of perturbation in section \ref{sec:3}.
\par

In this case, we observe certain non-linear density modulations at the two interfaces shared between the three components of the system. The density profiles of the three-components of the phase-separated BEC during its time evolution has been presented in the Fig. \ref{fig:11}. Analogous to the previous cases, the interfaces are affected at different time instances. Owing to the trigger, the condensate densities are modulated into exponentially evolving complicated non-linear patterns as shown in Fig. \ref{fig:11}.

\par
Although the quench dynamics for varying parameters can be focused in a separate study, this peculiar case of sudden quench suggests a certain way to induce instabilities in a multi-component system. Such studies can be regarded highly merely for their versatility.

\bibliographystyle{apsrev4-2}
\bibliography{RTI.bib}
\end{document}